\documentstyle[aps,prd,preprint,tighten,floats,epsf]{revtex}
\font\bdi=cmmib10 scaled\magstep1
\font\bsy=cmbsy10 scaled\magstep1
\def\bi#1{\mbox{\bdi #1\/}}
\def\bdot{\,\mbox{\bsy\char'001}\,}    
\newenvironment{problem}[1]%
 {\smallskip\noindent{\bf #1}.\quad\em}
 {\rm\smallskip}

\def\ci{{\rm i}}
\def\ce{{\rm e}}
\def\d{{\rm d}}
\def\halft{{\textstyle\frac{1}{2}}}
\def\kb{{\bi k}}
\def\xb{{\bi x}}
\def\kbh{\hat{\kb}}
\def\ybh{\hat{{\bi y}}}
\def\Vc{{\cal V}}
\def\Ac{{\cal A}}
\def\Kc{{\cal K}}
\def\threeq{{\textstyle\frac{3}{4}}}
\def\fiveq{{\textstyle\frac{5}{4}}}
\def\Pc{{\cal P}}
\def\Mc{{\cal M}}

\def\be{\begin{equation}}
\def\ee{\end{equation}}

\begin{document}
\preprint{VAND-TH-99-02}

\title{Inverting the Sachs--Wolfe Formula: an Inverse Problem
       Arising in Early-Universe Cosmology}

\author{A. Berera$^1$\thanks{E-mail:
bereraa@ctrvax.vanderbilt.edu, ab@ph.ed.ac.uk;
address after October 1, 1999, Department of Physics and
Astronomy, University of Edinburgh, Edinburgh EH9 3JZ, Scotland}
and P. A. Martin$^2$\thanks{E-mail:pamartin@mines.edu;
address after September 1, 1999, Mathematics Department, Colorado School of
Mines, Golden CO 80401, USA}}

\address{$^1$
Department of Physics and Astronomy,
Vanderbilt University, Nashville, TN 37235, U.S.A. \\
$^2$ Department of Mathematics, University of Manchester, 
Manchester M13 9PL, England}

\maketitle

\begin{center}
26th January 1999. Revised 1st June 1999.
\end{center}

\begin{abstract}
The (ordinary) Sachs--Wolfe effect relates primordial matter
perturbations to the temperature variations $\delta T/T$
in the cosmic microwave background radiation; $\delta T/T$ can
be observed in all directions around us. A standard but idealised
model of this effect leads to
an infinite set of moment-like equations: the
integral of $P(k)\,j_\ell^2(ky)$ with respect to $k$ ($0<k<\infty$)
is equal to a given constant, $C_\ell$, for $\ell=0,1,2,\ldots$.
Here, $P$ is the power spectrum of the primordial density variations,
$j_\ell$ is a spherical Bessel function and $y$ is a positive constant.
It is shown how to solve these equations exactly
for~$P(k)$. The same solution can be recovered, in principle, if the
first~$m$ equations are discarded. Comparisons with classical moment
problems (where $j_\ell^2(ky)$ is replaced by $k^\ell$) are made.
\end{abstract}

\medskip

In Press {\it Inverse Problems} 1999

\eject

\section{Introduction}\label{intro}

Some might say that the ultimate inverse problem is to understand the
origin of structure in the Universe using information that is
currently available: it is the central problem in early-Universe
cosmology~\cite{peebook,kotu}. One of the available cosmological
observables is the cosmic microwave background radiation (CMBR).
It is known that
the temperature of the CMBR is remarkably uniform in its spatial
variation, with $\delta T/T\simeq 10^{-5}$--$10^{-4}$ \cite{smootea},
\cite[\S 1.5]{kotu},
implying that the expansion of the Universe is largely isotropic.
Conversely, inhomogeneities in the density of the Universe lead to
temperature anisotropies, so that these can be used as a sensitive
test of theories of structure formation~\cite{WSS94}.

In these theories, the primary unknown function is $P(k)$, the power
spectrum of the primordial density fluctuations. Under certain
assumptions, it can be shown that $P(k)$ satisfies the following set
of equations:
\be
  \int_0^\infty k^{-2}\,P(k)\,j_\ell^2(ky)\,dk =C_\ell,\qquad
  \ell=0,1,2,\ldots.           \label{GOVEQ}
\ee
Here, $y$ is a given positive constant, $j_\ell$ is a spherical Bessel
function and the constants $C_\ell$ are given.

Experiments have provided estimates for a finite number of the
$C_\ell$'s. How can these be used to recover~$P(k)$? This is a major
question, but it is {\it not\/} our main concern here. We are
interested, first, in an idealised problem: given an exact knowledge
of all the $C_\ell$'s, reconstruct~$P$. We show that this inverse
problem can be solved exactly. This is a new result. Furthermore, we
show that the same solution can be recovered if the first few (in
fact, any finite number) of the $C_\ell$'s are unknown.

What use is an exact inversion formula for an idealised problem?
First, it reveals the ill-posed nature of the problem.\footnote{For
the definition of an ill-posed problem, see, for
example~\cite[ch~4]{CK92}.}  Thus, $P(k)$ is
found to be given by the Fourier sine transform of a certain function
$g_0(\lambda)$, which is defined by an infinite series with a finite
interval of convergence. This means that techniques of analytic
continuation are required, a process that can be very difficult
numerically~\cite{Henrici}
and one that gives the problem its ill-posed character.
Second, our inversion formula reveals some of the
analytic structure. For example, the low-$k$ behaviour of $P(k)$ is
intimately related to the asymptotic behaviour of $g_0(\lambda)$ as
$\lambda\to\infty$. Third, exact results can be used to test numerical
algorithms designed for the finite-data problem.

The plan of the paper is as follows. In the next section, we sketch a
derivation of the governing equations~(\ref{GOVEQ}). Careful
derivations can be found in the literature; references are given. Our
aim is to motivate the study of~(\ref{GOVEQ}) in a way that is
accessible to non-specialists. Thus, we limit our discussion to
perhaps the simplest model of the underlying physics.
In section~\ref{FORMSECT}, we formulate two moment-like problems,
called the Basic Problem and the Reduced-data Problem in which we are
given $C_\ell$ for $\ell\geq 0$ and $\ell\geq\ell_0$, respectively,
where $\ell_0$ is any fixed positive integer.
Apart from the system~(\ref{GOVEQ}), we also consider
(in section~\ref{sectfour}) the
analogous classical moment problem where $j_\ell^2(ky)$ is replaced
by~$k^\ell$. For both cases, we apply a general method described in
section~\ref{GENSECT}. This replaces the infinite system of
moment-like equations by a single integral equation. Application to
the astrophysical problem is made in
sections~\ref{sectfive}--\ref{secteight}.

\section{The governing equations}\label{gesect}

There are several physical causes of temperature variations in the
CMBR (Kolb and Turner list five~\cite[p~383]{kotu}). For large
angular scales, the dominant contribution comes from variations in the
gravitational potential; this is known as the {\it Sachs--Wolfe
effect\/}~\cite{SachsW}.
An elementary discussion of the relevant equations can be found
in~\cite[\S 9.6.2]{kotu}, \cite[\S 6.4]{padman} and
in~\cite{peebles,bunn,BFH98}, whereas advanced, detailed,
derivations that are up-to-date with the present state of computation 
can be found in~\cite{ref2f} and~\cite{ref2g}.
A brief derivation is sketched here; for
a rigorous derivation, involving time-dependent processes within
General Relativity, see the cited books and papers.

Consider the standard case of a flat universe ($\Omega_0=1$) with
zero cosmological constant ($\Lambda=0$). The observed temperature
fluctuations of the CMBR in the
direction of the unit vector~$\ybh$ are
\[
  \frac{\delta T(\ybh)}{T}=\frac{T(\ybh)-T_0}{T_0},
\]
where $T_0$ is the mean temperature.
The Sachs--Wolfe effect relates the observed $\delta T/T$ to the
primordial density fluctuations in the early Universe at a time
when light and matter decoupled. For the microwave radiation, the
whole Universe at this time is referred to as the {\it last scattering
surface\/}~\cite[p~74]{kotu}. Since the time of last scattering to the
present, the distance travelled by light is $y=2c/H_0$, where
$c$ is the speed of light and $H_0$ is the Hubble constant.
A present-day observer at a given point will receive
light from a spherical shell of present-day radius $y$ on the last
scattering surface. Thus, it is natural to expand on that shell
in terms of spherical harmonics $Y_n^m(\ybh)$,
\be
  \frac{\delta T (\ybh)}{T}
  = \sum_{\ell, m} a_{\ell m}(y) Y_\ell^m(\ybh),   \label{deltYY}
\ee
where $a_{\ell m}$ are dimensionless expansion coefficients.

We can also give a Fourier representation for $\delta T/T$,
\be
  \frac{\delta T (\ybh)}{T}
     = \Vc\int \Theta(\kb) \, \exp{(\ci y \kb \bdot \ybh)}\,\d\kb,
                                                   \label{delt}
\ee
where $\Theta(\kb)$ is a Fourier amplitude. The factor
$\Vc=V/(2\pi)^3$ where $V$ is a (large) volume, so that $\Theta$ is
dimensionless. (For simplicity, we have used a continuum description
(Fourier transform) here rather than a discrete description (Fourier
series) followed by an appropriate limit.)

We can compare~(\ref{deltYY}) and~(\ref{delt}), using the plane-wave
expansion
\[
  \exp{(\ci y\kb\bdot\ybh)}=
  4\pi\sum_{\ell, m} i^l\,j_l(ky)
  \,Y_\ell^m(\ybh)\,\overline{Y_\ell^m(\kbh)},
\]
where $j_\ell$ is a spherical Bessel function, $\kb=k\kbh$, $k=|\kb|$
and the overbar denotes complex conjugation. We find that
\[
  a_{\ell m}(y)= 4\pi i^\ell \Vc\int \Theta(\kb)\,
         j_\ell(ky)\,\overline{Y_\ell^m(\kbh)}\,\d\kb,
\]
whence use of the addition theorem for Legendre polynomials gives
\be
 \sum_m |a_{\ell m}|^2= 4\pi(2\ell+1) \,\Vc^2 \int\!\!\!\int
    \Theta(\kb)\,\overline{\Theta(\kb')}\, j_\ell(ky)\,
      j_\ell(k'y)\,P_\ell(\kbh\bdot\kbh')\,\d\kb\,\d\kb'.  
\label{SUMAA}
\ee

The Sachs--Wolfe effect relates $\delta T/T$, through physics on the
last scattering surface, to the primordial density fluctuations at
time of last scattering, $\delta\rho/\rho$. The latter quantity has a
Fourier representation similar to~(\ref{delt}), namely
\[
  \frac{\delta \rho (\xb)}{\rho}
     =\Vc \int \Delta(\kb) \, \exp{(\ci\kb\bdot\xb)}\,\d\kb.
\]
It is usual to assume that $\Delta(\kb)$ is an independent Gaussian
random variable of zero mean. Thus $\langle\Delta(\kb)\rangle=0$
and
\[
  \left\langle\Delta(\kb)\,\overline{\Delta(\kb')}\right\rangle =
   \Vc^{-1}  P(k)\,\delta^3(\kb-\kb'),
\]
where $P(k)$ is the {\it power spectrum\/} of the density
fluctuations. Note that
\be
   P(k)=\Vc\int \left\langle\Delta(\kb)\,
        \overline{\Delta(\kb')}\right\rangle\, \d\kb'. \label{Pdefn}
\ee
Then, the Sachs--Wolfe analysis shows that (the Fourier transforms of)
$\delta T/T$ and $\delta\rho/\rho$ are related by
\be
   \Delta(\kb)= \Ac k^2\,\Theta(\kb),         \label{query}
\ee
where $\Ac$ is a real constant with the dimensions of area. (Roughly
speaking, the temperature fluctuations are proportional to the
perturbed gravitational potential, $\Phi$ say, and $\nabla^2\Phi$ is
proportional to the density fluctuations; from~\cite{peebles}, we have
$\Ac=-2c^2/H_0^2=-\halft y^2$.) So, if we take an ensemble average
of~(\ref{SUMAA}) and define rotationally-invariant dimensionless
coefficients $C_\ell$ by
\[
  C_\ell \equiv \frac{1}{2\ell+1}
  \sum_m \left\langle |a_{\ell m}|^2\right\rangle,
\]
we find that
\begin{eqnarray}
  C_\ell &=& \frac{4\pi \Vc}{\Ac^2} \int k^{-4}P(k)\, j_\ell^2(ky)\,\d\kb
                                          \nonumber\\
     &=& (4\pi/\Ac)^2 \Vc\int_0^\infty k^{-2}P(k)\, j_\ell^2(ky)\,\d k.
                                          \label{CPjj}
\end{eqnarray}
The set of coefficients~$\{C_\ell\}$ is known as the {\it angular
power spectrum}.\footnote{In General Relativity, a density fluctuation
is not a
gauge-invariant quantity. In practical terms this means density
fluctuations are not well-defined  observables.  For sub-Hubble scales,
the gauge dependence of density fluctuations is negligible, so that any
reasonable gauge choice is operationally acceptable.  However for
super-Hubble scales, the gauge dependence is acute.  Properly, the power
spectrum $P(k)$ in~(\ref{Pdefn}) must be defined with respect to
`gauge-invariant' quantities~\cite{bardeen,ks84} that in the sub-Hubble
scale regime are readily identified as density fluctuations.  This is  a
well-studied problem in the 
literature~\cite{mfb,ll,panek,sbb,gss91,magueijo,rp,hs95}.
For our purposes, however, these details are unimportant and we will be
content with our slightly imprecise description,~(\ref{Pdefn}).
What is important for our work is that the basic relation between
$P(k)$ and the angular power spectrum $C_{\ell}$ is
of the form~(\ref{CPjj}).}

The temperature fluctuations have been measured experimentally; for
example, this was done by the COBE satellite
(see \cite{smootea}, \cite{smoot} and references therein). This gives
an estimate of the angular power spectrum~\cite[\S 7]{bunn}: thus, we
have values of $C_\ell$ for a range of $\ell$-values. From these, we
would like to extract the power spectrum,~$P(k)$. This is a
moment-like problem. In section~\ref{sixsect}, we give an explicit
solution of an idealised problem, in which we know $C_\ell$ for
all~$\ell\geq 0$.

In practice, there are complications. First, the measured values
of $C_0$ and $C_1$, the `nuisance parameters'~\cite[p~175]{bunn}, are
unreliable. We show that $P(k)$ can be recovered, nevertheless, using
$C_\ell$ for all $\ell\geq\ell_0$, where $\ell_0$ is any fixed positive
integer.

Another complication is that we cannot actually measure $C_\ell$
because it is defined as an ensemble average: we always have a
relative uncertainty (or {\it cosmic variance\/})
of $(2\ell+1)^{-1/2}$, even if the experimental measurements are
perfect \cite{AW84}, \cite[p~354]{WSS94},
\cite[p~160]{bunn},~\cite[p~189]{smoot}.
Once inverted, these errors will lead to corresponding errors in the
calculated~$P(k)$. As the errors in $C_\ell$ decay as $\ell$ increases,
one possibility is to use $C_\ell$ for $\ell\geq\ell_0$, where $\ell_0$
is chosen sufficiently large, so as to recover~$P$. One could then
compute $C_\ell$ for $\ell<\ell_0$, and compare with the experimental
estimates, giving the possibility of cross-validation. However, we
defer discussions of computational aspects to a later paper.

Further complications come from the underlying physics. Thus,
the Sachs--Wolfe effect typically is decomposed into two additive
quantities, the `ordinary' and the `integrated' Sachs--Wolfe effects,
of which~(\ref{CPjj}) only represents the former. The methods
developed in this paper are only applicable to the ordinary
Sachs--Wolfe effect. In the idealized limit that the CMBR anisotropy is
composed only of super-Hubble scale, adiabatic, density perturbations
in a flat ($\Omega_0=1$) universe, the ordinary Sachs--Wolfe effect
dominates. This ideal limit has an approximate validity in inflationary
cosmology, which is a promising candidate for development
as a fundamental theory of the early universe; for reviews, see, for
example,~\cite[ch~8]{kotu} or~\cite[ch~10]{padman}.
The understanding
of the CMBR anisotropy is still a developing area of research, although
the general consensus is that there are substantial modifications to the
idealized regime of the ordinary Sachs--Wolfe effect. We will not
attempt to survey this vast body of research here, but the interested
reader can begin with a few exemplary 
papers~\cite{ks94,gss91,hssw95,hs95,bunn}.
It is safe to say that, when applied to real data, there would be
practical limitations to the methods developed in this paper.
Nevertheless, they offer a new and very different perspective on
the subject

\section{Formulation of two mathematical problems}\label{FORMSECT}

Let us begin by considering the following moment-like problem.

\begin{problem}{Basic Problem}
Given a set of numbers $\{C_n\}$ and a
set of functions $\{\psi_n(k)\}$, with $n=0,1,2,\ldots$, find a
function $f(k)$ (in a certain space) that satisfies
\be
  \int_0^\infty f(k)\,\psi_n(k)\,\d k=C_n,\quad \mbox{for
  $n=0,1,2,\ldots$.}                       \label{BASIC}
\ee
\end{problem}

We are mainly interested in one particular set $\{\psi_n\}$, given by
\be
  \psi_n(k)=j_n^2(ky),                 \label{SPHBESS}
\ee
where $y$ is fixed. For a
simpler model problem, we shall also consider the choice
\be
  \psi_n(k)=k^n,                        \label{MOMENT}
\ee
which gives rise to a classical (Stieltjes) moment
problem~\cite{Shohat,Akhiezer}.

We are also interested in the following related problem
where the first~$m$ of~(\ref{BASIC}) are omitted.

\begin{problem}{Reduced-data Problem}
Given a set of numbers $\{C_n\}$
and a set of functions $\{\psi_n(k)\}$, find a function $f_m(k)$
that satisfies
\be
  \int_0^\infty f_m(k)\,\psi_n(k)\,\d k=C_n,\quad \mbox{for
  $n\geq m\geq 0$.}                       \label{REDUCED}
\ee
Here, $m$ is fixed and $f_0\equiv f$.
\end{problem}

It is important to note that we seek $f_m$ in the {\it same\/} space
as~$f$. We are also using the same numbers $\{C_n\}$ and the same
functions $\{\psi_n\}$ (for $n\geq m$) as in the Basic Problem: thus,
$C_n$ and $\psi_n$ are independent of~$m$.

As we saw in section~\ref{gesect}, the Reduced-data Problem, with
$\psi_n(k)=j_n^2(ky)$, arises in connection with the Sachs--Wolfe
effect. In that context, the basic unknown function is the power
spectrum $P(k)=k^2f(k)$, apart from a constant factor;
see~(\ref{CPjj}).

\section{A general method}\label{GENSECT} 

A general method for treating moment-like problems is to replace
them by an integral equation. Thus, choose a second set of functions
$\{\phi_n(\lambda)\}$, multiply the $n$-th equation of~(\ref{REDUCED})
by~$\phi_n$ and sum over~$n$. This gives
\be
   \int_0^\infty f_m(k)\,\Kc_m(k,\lambda)\,\d k=g_m(\lambda),  \label{IEm}
\ee
where
\be
   \Kc_m(k,\lambda)=\sum_{n=m}^\infty \psi_n(k)\,\phi_n(\lambda)
                                                        \label{Kmsum}
\ee
and
\be
   g_m(\lambda)=\sum_{n=m}^\infty C_n\,\phi_n(\lambda).  \label{gmsum}
\ee
To be effective, one has to be able to solve the integral
equation~(\ref{IEm}),
perhaps by recognising the left-hand side as a known integral
transform; implicitly, this requires that the sum defining the kernel
$\Kc_m$ can be evaluated in closed form. To make the method rigorous,
one has to show that the sums in~(\ref{Kmsum}) and~(\ref{gmsum}) are
convergent, and that the implied interchange of summation and
integration is justified. Note that, at this stage, $\lambda$ is
unspecified.

We will use this method below for the two sets $\{\psi_n\}$ given
by~(\ref{MOMENT}) and~(\ref{SPHBESS}).
We remark that a similar method was used in~\cite{PAM80,PAM82} to
analyse the so-called {\it null-field equations}, a system of
moment-like equations (involving Hankel functions) that can be used to
solve the boundary-value problems for acoustic scattering by bounded
obstacles.

\section{Moment problems}\label{sectfour} 

We start with $\psi_n(k)=k^n$, giving the following simple Reduced-data
Problem.

\begin{problem}{Moment Problem~$\Mc_m$}
Find a function $f_m(k)$ that satisfies
\be
  \int_0^\infty f_m(k)\,k^n\,\d k=C_n\quad \mbox{for all $n\geq
  m\geq 0$.}                       \label{MmEQN}
\ee
We assume that $f_m$ is a smooth function that has a Laplace
transform.
\end{problem}

We can rewrite~(\ref{MmEQN}) as
\[
  \int_0^\infty \{k^m\,f_m(k)\}\,k^n\,\d k=C_{n+m},\quad \mbox{for
  $n=0,1,2,\ldots$.}
\]
Multiply by $\phi_n(\lambda)=(-\lambda)^n/n!$ and sum over~$n$ to give
\be
    \int_0^\infty \{k^m\,f_m(k)\}\,\ce^{-\lambda k}\,\d k=g_m(\lambda),
                                         \label{kmfmLT}
\ee
where
\be
  g_m(\lambda)=\sum_{n=0}^\infty C_{n+m} \frac{(-\lambda)^n}{n!}.
                                                    \label{gmDEFN}
\ee

Equation~(\ref{kmfmLT}) gives the Laplace transform of $k^mf_m(k)$, so
that we can invert to obtain~$f_m$ itself. This shows that $f_m$ is
obtainable uniquely from the given coefficients $\{C_n\}$ with~$n\geq
m$.

In the derivation above, we assumed that the series~(\ref{gmDEFN}) is
convergent. In fact, we only require convergence for
$|\lambda|<\lambda_0$, say, and then define $g_m(\lambda)$ for
larger~$|\lambda|$ by analytic continuation. The assumption that
$f_m(k)$ has a Laplace transform is enough to ensure that $k^mf_m(k)$
also has a Laplace transform, this being what is actually
required in the derivation.

The connection between classical moment problems and the Laplace
transform is well known; see, for example, \cite[p~97]{Shohat}
and~\cite[p~230]{feller}. For a connection with a Hankel transform
(take $\phi_n(\lambda)=(-\lambda)^n/(n!)^2$), see~\cite[p~96]{Shohat}.

How are $f_m$ and $f_0\equiv f$ related, for $m>0$?
    From~(\ref{gmDEFN}), we have
\begin{eqnarray*}
  g_m(\lambda)&=&\sum_{n=m}^\infty C_n \frac{(-\lambda)^{n-m}}{(n-m)!}
     =\left(-\frac{\d}{\d\lambda}\right)^m\sum_{n=m}^\infty C_n
     \frac{(-\lambda)^n}{n!}\\
     &=&\left(-\frac{\d}{\d\lambda}\right)^m\sum_{n=0}^\infty C_n
     \frac{(-\lambda)^n}{n!}
     =\left(-\frac{\d}{\d\lambda}\right)^m g_0(\lambda).
\end{eqnarray*}
We also have
\[
  \int_0^\infty \{k^m\,f_m(k)\}\,\ce^{-\lambda k}\,\d k=
  \left(-\frac{\d}{\d\lambda}\right)^m
  \int_0^\infty f_m(k)\,\ce^{-\lambda k}\,\d k.
\]
So, integrating $m$ times gives
\be
 \int_0^\infty f_m(k)\,\ce^{-\lambda k}\,\d k = g_0(\lambda)
           +p_m(\lambda),                      \label{LT}
\ee
where $p_0\equiv 0$ and
\[
  p_m(\lambda)=\sum_{\ell=0}^{m-1} a_\ell \lambda^\ell
\]
is an arbitrary polynomial of degree~$(m-1)$. But the left-hand side
of~(\ref{LT}) is a Laplace transform, and so it must vanish as
$|\lambda|\to\infty$ (in a right-hand half-plane). $g_0(\lambda)$ has
the same property, as it is the Laplace transform of~$f_0$. Hence, the
polynomial $p_m(\lambda)$ must be absent:
\[
 \int_0^\infty f_m(k)\,\ce^{-\lambda k}\,\d k = g_0(\lambda)
  = \int_0^\infty f_0(k)\,\ce^{-\lambda k}\,\d k.
\]
Thus, $f_m(k)=f_0(k)$ for all~$m$. This means that if we know, {\it a
priori}, that $f_m$ and $f_0$ are both in the {\it same\/} space
(here, we assumed that they both have Laplace transforms), then they
are equal: deleting the first $m$ moments $C_n$ does not represent a
loss of information. Similar comparisons can be made between $f_m$ and
$f_{m'}$ with $m\neq m'$.

For a simple example, take $C_n=n!$ for $n\geq 0$, whence
\[
  g_0(\lambda)=\sum_{n=0}^\infty (-\lambda)^n =(1+\lambda)^{-1}
\]
for $|\lambda|<1$. Note that we can define $g_0(\lambda)$ for all
$\lambda\neq -1$ by analytic continuation. By inspection, we have
$f(k)=\ce^{-k}$.

\section{The Sachs--Wolfe effect}\label{sectfive} 

Here, we assume that $\psi_n(k)=j_n^2(ky)$, where $j_n$ is a spherical
Bessel function and $y$ is fixed. Thus, we consider the following
problem.

\begin{problem}{Sachs--Wolfe Problem~$\Pc_m$}
Find a function $f_m(k)$ that satisfies
\be
  \int_0^\infty f_m(k)\,j_n^2(ky)\,\d k=C_n\quad \mbox{for all $n\geq
  m\geq 0$.}                       \label{mEQN}
\ee
Here, $m$ is fixed, and the constants $C_n$ are given.
\end{problem}

What conditions on $f_m$ are appropriate? Let us assume that
\[
  f_m(k)\sim \left\{
  \begin{array}{ll}
  k^\alpha &\mbox{as $k\to 0$,}\\
  k^{1-\beta} &\mbox{as $k\to\infty$.}
  \end{array}\right.
\]
Then, considering the integral on the left-hand side of~(\ref{mEQN}),
we see that we need
\[
            2m+\alpha+1>0\quad\mbox{for convergence at $k=0$}
\]
and
\[
            \beta>0\quad\mbox{for convergence at $k=\infty$.}
\]
Note that convergence at $k=0$ depends on the smallest value of~$n$
used, which is~$m$. However, we want conditions that do not depend
on~$m$. So, given that the Basic Problem is~$\Pc_0$, minimal
conditions are
\be
  \mbox{$\alpha>-1$ and $\beta>0$.}        \label{CONCON}
\ee

Arguments based on causality imply that $P(k)=O(k^4)$ as $k\to
0$ \cite{AT86,RW96}, where $P(k)=k^2f(k)$. Thus, the desired physical
solution should have~$\alpha\geq 2$. This has led to studies of model
power spectra of the form
\be
   f(k)=\frac{(ky)^\alpha}{1+(k/k_0)^{\alpha+\beta-1}}, \label{modelf}
\ee
where $\alpha\geq 2$, $\beta>0$ and $k_0$ is a constant~\cite{BFH98}.

Next, let us consider an exact solution~\cite[formula 6.574.2]{GR80}:
\be
 \int_0^\infty k^{1-\beta}\,j_n^2(ky)\,\d k=
 \frac{\pi y^{\beta-2}\Gamma(\beta)\,\Gamma(n-\halft\beta+1)}
 {2^{\beta+1}\{\Gamma(\halft(\beta+1))\}^2\,\Gamma(n+\halft\beta+1)}.
                         \label{EXACTjn}
\ee
This holds for $0<\beta<2n+2$. Note that the right-hand side is
$O(n^{-\beta})$ as $n \to\infty$.

This exact solution throws some light on the next question:
how does the integral on the left-hand side of~(\ref{mEQN}) behave as
$n\to\infty$? If we split the range of integration at $k=1$, say, we
can easily see that the asymptotic contribution from integrating over
$0\leq k\leq 1$ is exponentially small. The
dominant contribution comes from integrating over $k>1$, so that the
large-$k$ behaviour of $f_m(k)$ can be used. Comparing with the exact
solution given above shows that
\[
       C_n=O(n^{-\beta})\quad\mbox{as $n\to\infty$.}
\]
Thus, there is a link between the large-$n$ behaviour of the (given)
coefficients $C_n$ and the large-$k$ behaviour of~$f_m(k)$.

Another exact solution is~\cite[formula 6.577.1]{GR80}:
\be
    \int_0^\infty \frac{k^2}{k^2+k_0^2}\,j_n^2(ky)\,\d k=
 \frac{\pi}{2y}\, I_{n+1/2}(k_0y)\,K_{n+1/2}(k_0y), \label{EXACTrat}
\ee
where $I_{n+1/2}$ and $K_{n+1/2}$ are modified Bessel functions and
$k_0$ is a positive constant. Note that the right-hand side is
$O(n^{-1})$ as $n \to\infty$. This formula gives the angular power
spectrum for the model~(\ref{modelf}), exactly, when $\alpha=2$
and~$\beta=1$. Further formulae can be obtained by differentiation
of~(\ref{EXACTrat}) with respect to~$k_0$.

\section{The Basic Problem~$\Pc_0$}\label{sixsect} 

We are going to apply the general method of section~\ref{GENSECT} to
problem~$\Pc_0$, which we restate here.

\begin{problem}{Problem~$\Pc_0$}
Find a function $f(k)$ that satisfies
\be
  \int_0^\infty f(k)\,j_n^2(ky)\,\d k=C_n\quad \mbox{for
  $n=0,1,2,\ldots$.}                       \label{P0EQN}
\ee
\end{problem}

We have to choose the set $\{\phi_n(\lambda)\}$.
There are very few known sums involving products of Bessel functions.
   From Abramowitz \& Stegun~\cite[formula 10.1.45]{AS65}, we have
\be
  \sum_{n=0}^\infty (2n+1)\,P_n(\cos\theta)\,j_n^2(ky)=
  \frac{\sin{\{ky\sqrt{2-2\cos\theta}\}}}{ky\sqrt{2-2\cos\theta}},
                                    \label{AS45}
\ee
where $P_n$ is a Legendre polynomial and~$\theta$ is unrestricted.
If we define~$\lambda$ by
\be
  \lambda=y\sqrt{2-2\cos\theta}=2y\sin{\halft\theta}, \label{LAMDEF}
\ee
we can rewrite~(\ref{AS45}) as
\be
   \Kc_0(k,\lambda)=\frac{\sin{k\lambda}}{k\lambda}, \label{K0SINK}
\ee
where $\Kc_m$ is defined by~(\ref{Kmsum}) and
\be
  \phi_n(\lambda)=(2n+1)\,P_n\left(1-\halft(\lambda/y)^2\right).
                                             \label{PHINP}
\ee
Note that, for convergence of the integral in~(\ref{IEm}), we require
that $\lambda$ is {\it real}. Moreover, we can obtain the
formula~(\ref{K0SINK}) for all $\lambda>0$ if we allow
$\theta=\theta_R+i\theta_I$ to be complex; specifically, we can suppose
that $\theta$ lies on the L-shaped contour given by
\[
  \mbox{$\{0<\theta_R\leq\pi$ with $\theta_I=0\}$ and
        $\{0\leq\theta_I<\infty$ with $\theta_R=\pi\}$.}
\]

So, multiplying (\ref{P0EQN}) by $(2n+1)\,P_n(\cos\theta)$ and summing
over~$n$ gives
\be
   \int_0^\infty k^{-1}f(k)\,\sin{\lambda k}\,\d k=
   \lambda\,g_0(\lambda),   \quad\mbox{for $\lambda>0$,} \label{FSTf}
\ee
where
\be
  g_0(\lambda)= \sum_{n=0}^\infty (2n+1)\,C_n\,
  P_n\left(1-\halft(\lambda/y)^2\right).      \label{g0CP}
\ee
This reduces the determination of $f(k)$ to the inversion of a Fourier
sine transform:
\be
  f(k)=\frac{2k}{\pi} \int_0^\infty
  \lambda\,g_0(\lambda)\,\sin{\lambda k}\,\d\lambda. \label{fksine}
\ee
The derivation assumes that $k^{-1}f(k)$ has a Fourier sine transform,
which is consistent with the convergence conditions~(\ref{CONCON}). It
also assumes that the series~(\ref{g0CP}) is convergent for some
values of~$\lambda$, and that $g_0(\lambda)$ can be defined for other
values of~$\lambda$ by analytic continuation.

For an example, take $C_n$ to be defined by the right-hand side
of~(\ref{EXACTjn}) for $n\geq 0$ with $0<\beta<2$. In order to
calculate $g_0(\lambda)$, defined by~(\ref{g0CP}), we use the
following expansion (see the appendix for a derivation):
\be
  (1-x)^{-\gamma}=\sum_{n=0}^\infty (2n+1)\,c_n\,P_n(x)\quad
  \mbox{for $|x|<1$,}              \label{LEGSUM}
\ee
where $\gamma<1$ and
\be
  c_n=2^{-\gamma}
  \frac{\Gamma(1-\gamma)}{\Gamma(\gamma)}
  \frac{\Gamma(n+\gamma)}{\Gamma(n-\gamma+2)}. \label{LEGCO}
\ee
Comparing with $C_n$, we set $\gamma=1-\halft\beta$ whence
\[
  c_n=(4/\pi) 2^{-\beta/2} y^{2-\beta} \Gamma(\beta)\,
  \sin{\left(\halft\pi\beta\right)}\,C_n.
\]
With $x=1-\halft(\lambda/y)^2$, we find that
\be
  g_0(\lambda)=\frac{\pi}{2}\frac{\lambda^{\beta-2}}
   {\Gamma(\beta)\,\sin{\left(\halft\pi\beta\right)}}. \label{g0EX}
\ee
Note that the series defining $g_0(\lambda)$ converges for $0<\lambda<2y$,
but we can use~(\ref{g0EX}) to define $g_0(\lambda)$ for
all~$\lambda>0$. Hence (\ref{fksine}) gives
\[
  f(k)=\frac{k}{\Gamma(\beta)\,\sin{\left(\halft\pi\beta\right)}}
   \int_0^\infty \lambda^{\beta-1} \sin{\lambda k}\,\d\lambda
   =k^{1-\beta},
\]
in agreement with~(\ref{EXACTjn}).

\section{The Reduced-data Problem~$\Pc_m$} 

For $m>0$, we consider the Reduced-data Problem~$\Pc_m$ (the
Sachs--Wolfe Problem) described in section~\ref{sectfive}.

Proceeding as for $\Pc_0$, we multiply
(\ref{mEQN}) by $(2n+1)\,P_n(\cos\theta)$ and sum over~$n$
from~$n=m$. This gives, using~(\ref{LAMDEF}),
\be
  \int_0^\infty f_m(k)\,\left\{
  \frac{\sin{\lambda k}}{\lambda k} -\sum_{n=0}^{m-1}\phi_n(\lambda)\,
  j_n^2(ky) \right\}\,\d k=   g_m(\lambda),  \label{eqn1.2}
\ee
where $g_m(\lambda)$ and $\phi_n(\lambda)$ are defined
by~(\ref{gmsum}) and~(\ref{PHINP}), respectively.

Next, multiply~(\ref{eqn1.2}) by~$\lambda$ and write as
\be
  \int_0^\infty k^{-1}\,f_m(k)\,\sin{\lambda k}\,\d k
  -\lambda\sum_{n=0}^{m-1}\phi_n(\lambda)
  \int_0^\infty f_m(k)\, j_n^2(ky) \,\d k=\lambda\, g_m(\lambda).
                               \label{EQNXX}
\ee
Noting that $P_n(x)$ is a polynomial in~$x$ of degree~$n$, we see that
the second term on the left-hand side of~(\ref{EQNXX}) is a polynomial
in~$\lambda$ of degree~$(2m-1)$. So, applying the differential
operator $\d^{2m}/\d\lambda^{2m}$ gives
\[
   \frac{\d^{2m}}{\d\lambda^{2m}}\int_0^\infty k^{-1}\, f_m(k)\,
   \sin{\lambda k}\,\d k=
   \frac{\d^{2m}}{\d\lambda^{2m}}\left( \lambda\,g_m(\lambda)\right)
  =\frac{\d^{2m}}{\d\lambda^{2m}}\left( \lambda\,g_0(\lambda)\right).
\]
Integrating $2m$ times then gives
\be
  \int_0^\infty k^{-1}\,f_m(k)\,\sin{\lambda k}\,\d k=
    \lambda\,g_0(\lambda) +p_{2m}(\lambda),  \label{FSTfm}
\ee
where $p_{2m}$ is an arbitrary polynomial of degree~$(2m-1)$.
But the left-hand side of~(\ref{FSTfm}) is the Fourier sine transform
of~$k^{-1}\,f_m(k)$. This is assumed to exist and, by the
Riemann--Lebesgue lemma, it must vanish as
$|\lambda|\to\infty$. $\lambda\,g_0(\lambda)$ has
the same property, as it is the Fourier sine transform of~$k^{-1}\,f$,
by~(\ref{FSTf}). Hence, the
polynomial $p_{2m}(\lambda)$ must be absent. It follows that
$f_m(k)=f(k)$ for all~$m$.

The solution given above is correct but somewhat deceptive, for in the
formulation of~$\Pc_m$ we are not given $C_n$ for $0\leq n<m$. Thus,
we cannot form the series~(\ref{g0EX}) defining $g_0(\lambda)$. So,
instead of~(\ref{FSTfm}), we obtain
\[
  \int_0^\infty k^{-1}\,f_m(k)\,\sin{\lambda k}\,\d k=
    \lambda\,g_m(\lambda) +p_{2m}(\lambda),
\]
where the polynomial $p_{2n}(\lambda)$ is to be determined by the
requirement that the right-hand side vanishes as $|\lambda|\to\infty$.
Thus (just as for $\Pc_0$), we have to effect the analytic
continuation of $g_m(\lambda)$. Numerical consequences remain to be
explored.

\section{Discussion}\label{secteight} 

We have given an explicit formula for solving the Basic
Problem~$\Pc_0$. For the application we have in mind, all the
constants $C_n$ are {\it non-negative\/}:
does it follow that the solution $f(k)$
is positive? In general, the answer is `no'. We show this by
giving an explicit counter-example. Let
\be
  f(k)=y\left\{ (ky/\pi)^{1/2}-(2/\pi)A\right\}, \label{CEX}
\ee
where $A$ is a constant to be chosen. Note that if $A>0$, $f(k)<0$ for
$k<4A^2/(\pi y)$. (The various factors in~(\ref{CEX}) are merely
inserted for algebraic convenience.) We observe that the~$f$ defined
by~(\ref{CEX}) is a linear combination of two of the exact
solutions~(\ref{EXACTjn}), corresponding to $\beta=\halft$ and
$\beta=1$. Thus,
\[
  C_n=\frac{\pi\,\Gamma(n+\threeq)}
   {2^{3/2} \left\{\Gamma(\threeq)\right\}^2 \,\Gamma(n+\fiveq)}
   -\frac{A}{2n+1}.
\]
We have to show that $C_n>0$ for all~$n\geq 0$; we do this using an
inductive argument. Straightforward calculation shows that
\[
 C_{n+1}=\frac{4n+3}{4n+5}\,C_n+\frac{4(n+1)}{(4n+5)(2n+1)(2n+3)}\,A.
\]
So, $C_n>0$ for all $n\geq 0$ provided $A>0$ and $C_0>0$. But
\[
  C_0=1-A,
\]
so we can take any $A$ with $0<A<1$.

\section*{Appendix}

Here, we derive the expansion~(\ref{LEGSUM}) with~(\ref{LEGCO}).
   From~(\ref{LEGSUM}), orthogonality gives
\begin{eqnarray*}
 c_n&=& \halft\int_{-1}^1 (1-x)^{-\gamma} P_n(x)\,\d x\\
  &=& \frac{(-1)^n}{2^{n+1} n!} \int_{-1}^1 (1-x)^{-\gamma}
       \frac{\d^n}{\d x^n} (1-x^2)^n\,\d x
\end{eqnarray*}
using Rodigues' formula. Integrating by parts $n$ times, using
\[
  \frac{\d^n}{\d x^n} (1-x)^{-\gamma}=
  \frac{\Gamma(n+\gamma)}{\Gamma(\gamma)} (1-x)^{-\gamma-n}
\]
and noting that the integrated terms vanish, gives
\[
  c_n=\frac{1}{2^{n+1} n!} \frac{\Gamma(n+\gamma)}{\Gamma(\gamma)}
  \int_{-1}^1 (1-x^2)^n (1-x)^{-\gamma-n}\,\d x.
\]
In the integral, put $x=1-2t$; it becomes
\[
  2^{n+1-\gamma}\int_0^1 t^{-\gamma} (1-t)^n\,dt
  =  2^{n+1-\gamma}\frac{n!\,\Gamma(1-\gamma)}{\Gamma(n-\gamma+2)},
\]
using the integral definition of the Beta function, and the
result~(\ref{LEGCO}) follows.

\section*{Acknowledgments}

Financial support was provided to AB by the U.~S.~Department of Energy.

\end{document}